# Blochnium-Based Josephson Junction Parametric Amplifiers: Superior Tunability and Linearity

A. Salmanogli[1], H. Zandi[2,4], and M. Akbari[3]
[1]Ankara Yildirim Beyazit University, Engineering Faculty, Electrical and Electronic Department, Ankara, Turkey
[2]Faculty of Electrical Engineering, K.N. Toosi University of Technology, Tehran, Iran
[3]Quantum Optics Lab, Departmant of Physics, Kharazmi University, Tehran, Iran
[4]Iranian Quantum Technologies Research Center (IQTEC), Tehran, Iran

**Abstract:**
The weak quantum signal amplification is an essential task in quantum computing. In this study, a recently introduced structure of Josephson junctions array called Blochnium (N series Quarton structure) is utilized as a parametric amplifier. We begin by theoretical deriving the system's Lagrangian, quantum Hamiltonian, and then analyze the dynamics using the quantum Langevin equation. By transforming these equations into the Fourier domain and employing the input-output formalism, leading metric indicators of the parametric amplifier become calculated. The new proposed design offers significant advantages over traditional designs due to its ability to manipulate nonlinearity. This premier feature enhances the compression point ($P_{1dB}$) of the amplifier dramatically, and also provides its tunability across a broad band. The enhanced linearity, essential for quantum applications, is achieved through effective nonlinearity management, which is theoretically derived. Also, the ability to sweep the C-band without significant spectral overlap is crucial for frequency multiplexing in scalable quantum systems. Simulation results show that Blochnium parametric amplifiers can reach to a signal gain around 25 dB with a compression point better than of -92 dBm. Therefore, our proposed parametric amplifier, with its superior degree of freedom, surpasses traditional designs like arrays of Josephson junctions, making it a highly promising candidate for advanced quantum computing applications.

**Keywords:** Josephson parametric amplifier, Blochnium, quantum Langevin equation, compression point

**Introduction:**
A Josephson Parametric Amplifier (JPA) is a crucial component in quantum circuit processing and quantum computing due to its ability to amplify weak quantum signals with minimal added noise. JPAs exploit the nonlinearity of Josephson junctions to achieve parametric amplification, making them highly effective at boosting signals without significantly degrading their quantum properties [1-9]. This is essential for reading out qubit states, where the signal is extremely weak and susceptible to multiple noise sources. The appropriate gain and low noise figure of JPAs ensure that the integrity of quantum information is preserved during amplification. Additionally, JPAs can operate over a broad range of frequencies and can be tuned in real-time, providing flexibility for various quantum systems [4-9].
However, in quantum applications, particularly at cryogenic temperatures, JPAs are often preferred over CMOS [10-12] and HEMT amplifiers [13-14] due to their unique ability to operate at the quantum limit of noise. JPAs are designed with superconducting Josephson junctions, which allow them to achieve ultra-low noise figures, typically below 0.5 dB. This characteristic is crucial in quantum systems, where preserving the quantum nature of signals is paramount. The noise performance of JPAs is superior to both CMOS and HEMT technologies, which, although they offer low noise figures at cryogenic temperatures, cannot match the quantum-limited performance of JPAs. This low noise is essential for accurately reading out quantum states and minimizing decoherence, making JPAs the preferred choice in highly sensitive quantum experiments and quantum computing systems. Moreover, JPAs operate at much lower temperatures (around 10 mK) than CMOS and HEMT amplifiers, which typically function at around 4.2 K. While CMOS and HEMT technologies offer advantages in terms of integration, scalability, and bandwidth, their limitations in noise performance, temperature compatibility, and power efficiency make them less ideal for sub-cryogenic (~ 10 mK) quantum applications [10-14]. JPAs, despite their narrower bandwidth and more challenging scalability, provide unparalleled noise performance and compatibility with quantum systems, making them the amplifier of choice in applications where maintaining quantum coherence and minimizing noise are of utmost importance [6-9].

There are diverse types of JPAs that can be utilized in different applications. Degenerate JPAs, can be phase-sensitive or phase-preserving amplifiers, which operate with two modes and provide higher gains compared to non-degenerate JPAs [5, 6]. This kind of JPA can amplify only one quadrature of the signal. Their structure includes a Josephson junction and a resonator with a carefully designed impedance environment. The gain is fine-tuned through the phase and amplitude of the pump signal. In non-degenerate JPAs [5, 6] however, the signal and idler modes are separated significantly. Therefore, the structures modes can be divided in terms of even and odd modes with different specs. The structure typically consists of at least two Josephson junctions embedded in a resonator. Gain control is achieved by tuning the pump power and frequency, while linearity is maintained by optimizing the junction parameters. In line with advancements in JPA engineering and the optimization of its technical features, novel designs like the Josephson Traveling-Wave Parametric Amplifier (JTWPA) have been introduced [5, 7-8]. These designs aim to enhance key characteristics of the JPA, offering broadband amplification and a high dynamic range, achieved through the use of a long chain of Josephson junctions. Their structure comprises a long transmission line with periodically spaced Josephson junctions. Gain control is managed by the length of the transmission line and the pump power, while linearity is enhanced by distributing the nonlinearity over many junctions, resulting in very low noise due to the distributed nature of the amplifier and reduced single-point nonlinearity. Therefore, JPA design trends have recently shifted towards enhancing the amplifier's linearity to safely amplify the fundamental harmonic [5-8]. In line with this, this work focuses on designing a specialized JPA to improve the amplifier's linearity. Therefore, we introduce a new structure as Blochnium [15-17] operating as a JPA for the first time based on our knowledge. We think that this unique structure gives some degree of freedom by which one can design a high performance JPA with far fewer elements. It represents an evolution in the design of Josephson junction-based qubits, specifically structured to exploit the phenomena of Bloch oscillations. This work shows that the special features of Blochnium JPA (BJPA) include its enhanced compression point ($P_{1dB}$) [18], which is a measure of the amplifier's linearity and power handling capability. Additionally, it can sweep a broad range of frequencies, such as the C-band (4 GHz - 8 GHz), without overlapping spectra; this feature is crucial for frequency multiplexing in scalable quantum systems, as it reduces the potential for quantum crosstalk and errors. The ability to sweep frequencies without spectral overlap further enhances their utility in scalable quantum architectures. These features make JPAs indispensable for advancing quantum technologies, offering superior performance compared to conventional amplifiers. In the following, we focus on the theoretical backgrounds and definition of the system, and its analyzing using quantum theory to fully understanding about its dynamics equation of motions.

## Theoretical backgrounds

*System definition*

In this work, we investigate the use of Blochnium structure as a JPA. Initially, we provide a comprehensive description of the JPA, as illustrated schematically in Fig. 1. The design incorporates N Quarton structures [15-17], each of them consists of M slave SQUIDs with the same Josephson energy of $E_{Js}$, and a master SQUID with Josephson energy $E_{Jm}$. One of the important elements (indeed it is a parasitic element) that can affect the performance of the system is $C_g$, which actually is hardly a controllable parameter in the quantum system depicted. It determines the equivalent impedance of the quantum system that should be matched with $Z_0$, which is the λ/4 resonator intrinsic impedance. Introducing any mismatching between the impedances deteriorate the coupling rate κ between the system and environment. Using quantum theory, we demonstrate that the total Hamiltonian of this structure significantly differs from that of Fluxonium and arrays of SQUIDs [6-7, 9]. This distinction is not only evident in the physical structure of Blochnium but also in its effective manipulation of JJ nonlinearity. In addition, we will show that BJPA exhibits unique features unattainable with other structures. In the following, we try to dive into the quantum mechanical aspects, theoretically deriving key JPA parameters such as gain and nonlinearity factors. Our analysis reveals the superior performance of Blochnium in terms of tunability and linearity, highlighting its potential for advanced quantum applications. This comprehensive approach underscores the distinct advantages of BJPA over traditional designs, making it a promising candidate for enhancing the efficiency and scalability of quantum systems.

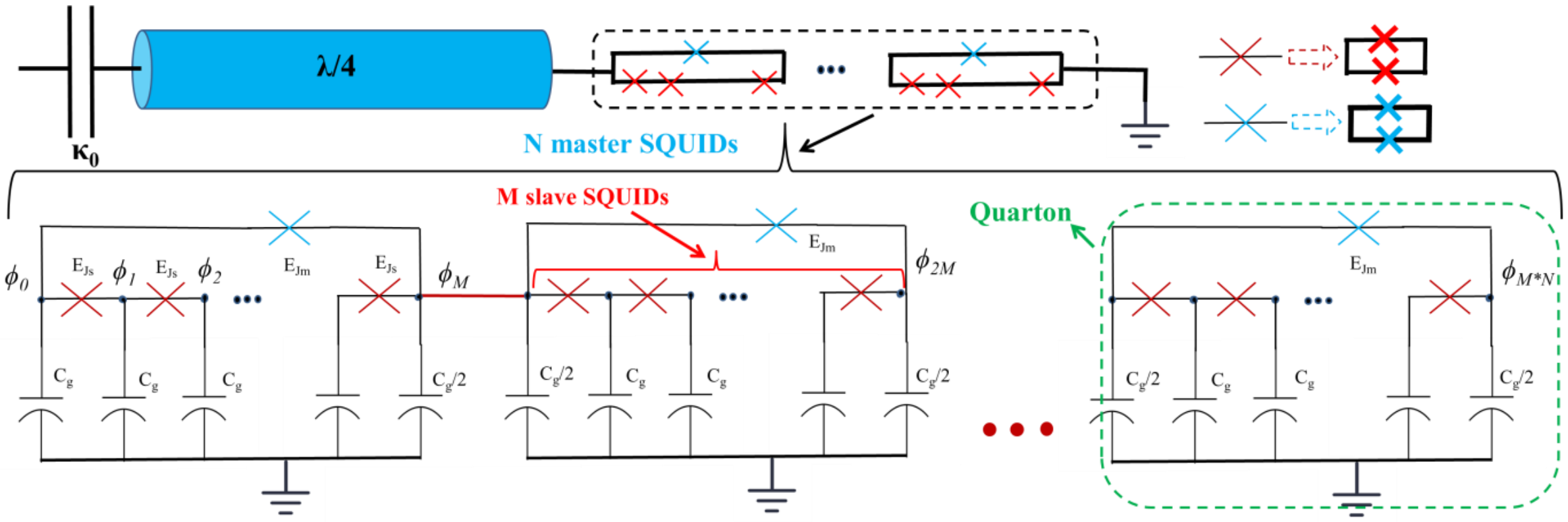

Fig. 1 The electrical schematic of a Blochnium structure (as a λ/4 resonator [6, 9]) containing N Quarton, meaning N master SQUIDs and each Quarton contains M slave SQUIDs.

*Analyzing of the system using Quantum theory*

To fully understanding about the structure proposed in Fig. 1, the approach begins with the theoretical derivation of the system's Lagrangian and total Hamiltonian, followed by an analysis of the dynamics using the quantum Langevin equation. The total Lagrangian [20] of a Blochnium shown in Fig. 1 is expressed as:

$$L_t = \frac{C_g}{2}\sum_{k=0}^{M*N}\dot{\phi}_k^{\,2} + \frac{C_{Js}}{2}\sum_{k=0}^{M*N-1}\left(\dot{\phi}_k - \dot{\phi}_{k+1}\right)^2 + \frac{C_{Jm}}{2}\sum_{k=0}^{N-1}\left(\dot{\phi}_{M*k} - \dot{\phi}_{M*(k+1)}\right)^2$$

$$-\frac{1}{2L_{Js}}\sum_{k=0}^{M*N-1}\left(\phi_k - \phi_{k+1}\right)^2 - \frac{1}{2L_{Jm}}\sum_{k=0}^{N-1}\left(\phi_{M*k} - \phi_{M*(k+1)}\right)^2 + M*N*E_{Js}\cos\left(\frac{\varphi}{M*N}\right) + N*E_{Jm}\cos\left(\frac{\varphi}{N}\right) \quad (1)$$

where $C_g$, ($C_{Js}$, $L_{Js}$, $E_{Js}$), ($C_{Jm}$, $L_{Jm}$, $E_{Jm}$), N, M, and $\phi$ are the parasitic capacitors grounded the SQUIDs, the salve's JJ capacitance inductance, and Josephson energy, the master's JJ capacitance, inductance, and Josephson energy, the number of master JJs, the number of slave SQUIDs in each B-cell, and flux as a quantum coordinate operator, respectively. To simplify the algebra, it is convenient to represent Eq. 1 through coupled equations. In this simplification approach, the last two terms in the Lagrangian are initially put aside, however they will be added while calculating the total Hamiltonian. Henceforth, one can express Eq. 1 as a compact form $L_t = \frac{1}{2}\dot{\phi}^T \hat{C} \dot{\phi} - \frac{1}{2}\phi^T \hat{L}^{-1} \phi$, where $\hat{C}$ and $\hat{L}^{-1}$ are tri-angular matrixes as:

$$\hat{C} = \begin{bmatrix} \underbrace{\begin{matrix} C_g+C_j+C_M & -C_j & 0 & 0 & \bullet & -C_M \\ -C_j & C_g+2C_j & -C_j & 0 & \bullet & 0 \\ 0 & -C_j & \bullet & -C_j & \bullet & 0 \\ 0 & 0 & -C_j & \bullet & \bullet & \bullet \\ \bullet & \bullet & \bullet & \bullet & C_g+2C_j & -C_j \\ -C_M & 0 & \bullet & 0 & -C_j & C_g+2C_j+2C_M \end{matrix}}_{\text{The first Quarton}} & \bullet\ \bullet & \begin{matrix} 0&0&0&\bullet&\bullet&0 \\ 0&0&0&\bullet&\bullet&0 \\ 0&0&0&\bullet&\bullet&0 \\ \bullet&\bullet&\bullet&\bullet&\bullet&0 \\ \bullet&\bullet&\bullet&\bullet&0&0 \\ 0&0&0&\bullet&\bullet&0 \end{matrix} \\ \bullet & \ddots & \bullet \\ \begin{matrix} 0&0&0&\bullet&\bullet&0 \\ 0&0&0&\bullet&\bullet&0 \\ 0&0&0&\bullet&\bullet&0 \\ \bullet&\bullet&\bullet&\bullet&\bullet&0 \\ \bullet&\bullet&\bullet&\bullet&0&0 \\ 0&0&0&\bullet&\bullet&0 \end{matrix} & \bullet\ \bullet & \overbrace{\begin{matrix} C_g+2C_j+2C_M & -C_j & 0 & 0 & \bullet & -C_M \\ -C_j & C_g+2C_j & -C_j & 0 & \bullet & 0 \\ 0 & -C_j & \bullet & -C_j & \bullet & 0 \\ 0 & 0 & -C_j & \bullet & \bullet & \bullet \\ \bullet & \bullet & 0 & \bullet & C_g+2C_j & -C_j \\ -C_M & 0 & \bullet & 0 & -C_j & C_g+2C_j+C_M \end{matrix}}^{\text{The last Quarton}} \end{bmatrix}$$

$$\hat{L}^{-1} = L^{-1} \begin{bmatrix} \begin{matrix} 2&-1&0&0&\bullet&-1 \\ -1&2&-1&0&0&0 \\ 0&-1&\bullet&-1&0&0 \\ 0&0&-1&\bullet&\bullet&\bullet \\ \bullet&0&0&\bullet&2&-1 \\ -1&0&\bullet&0&-1&4 \end{matrix} & \bullet\ \bullet & \begin{matrix} 0&0&0&\bullet&\bullet&0 \\ 0&0&0&\bullet&\bullet&0 \\ 0&0&0&\bullet&\bullet&0 \\ \bullet&\bullet&\bullet&\bullet&\bullet&0 \\ \bullet&\bullet&\bullet&\bullet&0&0 \\ 0&0&0&\bullet&\bullet&0 \end{matrix} \\ \bullet & \ddots & \bullet \\ \begin{matrix} 0&0&0&\bullet&\bullet&0 \\ 0&0&0&\bullet&\bullet&0 \\ 0&0&0&\bullet&\bullet&0 \\ \bullet&\bullet&\bullet&\bullet&\bullet&0 \\ \bullet&\bullet&\bullet&\bullet&0&0 \\ 0&0&0&\bullet&\bullet&0 \end{matrix} & \bullet\ \bullet & \begin{matrix} 4&-1&0&0&\bullet&-1 \\ -1&2&-1&0&\bullet&0 \\ 0&-1&\bullet&-1&0&0 \\ 0&0&-1&\bullet&\bullet&\bullet \\ \bullet&0&0&\bullet&2&-1 \\ -1&0&\bullet&0&-1&2 \end{matrix} \end{bmatrix} \tag{2}$$

Using the matrices for capacitors and inductors presented in Eq. 2, we can define the matrix $\Omega^2 = \hat{C}^{-1}\hat{L}^{-1}$ [6, 9] to calculate the eigenvalues (dominant frequencies) and eigenvectors ($\Psi_i$ is the wave profile of each mode) of the quantum system under discussion. Using these equations, one can calculate the effective capacitance and inductance related to the structure. The effective capacitance and inductance of the circuits can be determined using the eigenvalues and eigenvectors for each mode as follows: $C_{eff} = \overrightarrow{\Psi_i}^T \hat{C} \overrightarrow{\Psi_i}$ and $L_{eff}{}^{-1} = \overrightarrow{\Psi_i}^T \hat{L}^{-1} \overrightarrow{\Psi_i}$. This allows us to map a λ/4 resonator to an equivalent LC circuit [6]. Consequently, the Blochnium structure designed in this work is modeled as a series combination of an effective inductors and capacitors, along with nonlinear elements due to the terms separated in Eq. 1. Analyzing this simplified circuit model enables us to determine much more easily and effectively the quantum circuit impedance $Z_{eff} = \sqrt{L_{eff}/C_{eff}}$ and subsequently obtain the coupling rate of the quantum system to the environment $\kappa_{eff} = \omega_{eff}/Q_{eff}$, where $Q_{eff}$ is the circuit quality factor related to mismatching. Therefore, the total Hamiltonian [20] of the effective circuit can be summarized as:

$$H_t = \hbar\omega_{eff} a^+ a - M*N*E_{Js}\cos\left(\frac{\varphi}{M*N}\right) - N*E_{Jm}\cos\left(\frac{\varphi}{N}\right) \tag{3}$$

where a and $a^+$ are the annihilation and creation operators, respectively. In the derivation of the total Hamiltonian it is supposed that the circuit contains a simple LC oscillators and a nonlinear element. One can simply represent the second and third term in the terms of the ladder operators using the Taylor's expansion of the cosine function [6, 7] up to the forth order as:

$$H_t = \hbar\omega_{eff} a^+ a - M*N*E_{Js}\left(\frac{\varphi^4}{4!M^4*N^4}\right) - N*E_{Jm}\left(\frac{\varphi^4}{4!N^4}\right) \tag{4}$$

It should be noted that the second term of the Taylor's series of the cosine expansion was absorbed by oscillatory function earlier. Finally, using the quantization for the phase expressed in the equation, and with the assumption that the $E_{Jm} = \alpha_c^* E_{Js}$, and also the frequency resonance of each Quarton in the structure should be constant by considering the later assumption, the total Hamiltonian can be introduced as:

$$H_t = \hbar\omega_{eff} a^+ a - \frac{E_c}{6N}\left(\frac{1}{M} - \alpha_c^*\right) a^{+2} a^2 \tag{5}$$

In this equation, derived nonlinearity terms are clearly dependent on the number of Quartons in the circuit, number of SQUIDs in each Quarton, and also the number ratio of master and slave JJs. Having the complete Hamiltonian calculated, one can analyze the system dynamics by quantum Langevin equation [18-22] and finally achieve the signal gain and other related parameters. The dynamics equation of motion for this system is examined (quantum Langevin equation) [20] as:

$$\dot{a} = -i\omega_{eff} a - iKa^+ aa - \frac{\kappa}{2} a + \sqrt{\kappa} a_{in} \tag{6}$$

In this equation, $K = -E_c/6\hbar N\left(1/M - \alpha_c^*\right)$, $\alpha_c^* \equiv \alpha_c/M$, where $\alpha_c$ is the nonlinearity factor, and "a" determines the intra-cavity signals expressed as $a = \alpha + \delta a$, where $\alpha$ and $\delta a$ are the DC point and quantum signal fluctuation around the DC point, respectively. The steady-state solution (replacing $a = \alpha$ and $a_{in} = \alpha_{in}$) [19] for a coherent pump $\alpha = |\alpha|\exp(i\varphi)$ after some algebra's simplification is introduced as:

$$\left[\frac{1}{4} + \left(\frac{\omega_p - \omega_{eff}}{\kappa}\right)^2\right]|\alpha|^2 - \frac{(\omega_p - \omega_{eff})K}{\kappa^2}|\alpha|^4 + \left(\frac{K}{\kappa}\right)^2|\alpha|^6 - \frac{K}{\kappa^2}|\alpha_{in}|^2 = 0$$

$$\xrightarrow{n=\frac{|\alpha|^2}{|\alpha_{in}|^2},\ \delta=\frac{\omega_p-\omega_{eff}}{\kappa},\ \zeta=\left(\frac{K}{\kappa}\right)|\alpha_{in}|^2,\ \hat{\alpha}_{in}=\alpha_{in}\frac{1}{\sqrt{\kappa}}} \left[\frac{1}{4} + \delta^2\right]n - 2\delta\zeta n^2 + \zeta^2 n^3 - 1 = 0 \tag{7}$$

Therefore, using Eq. 7, the normalized intra-cavity average number of photons (n) created by the pump effect can be calculated as the roots of the equation. In Eq. 7, δ and ζ represent the normalized pump detuning and the relative strength of the nonlinearity in the presence of the pump field, respectively. The results of the simulation are illustrated in Fig. 2a, showing how the intra-cavity average number of photons varies with changes in the relative strength of the nonlinearity. It is evident from the figure that increasing the nonlinearity can lead to bifurcation [7]. It is important to distinguish between K and ζ, where K represents the nonlinearity associated with the JJ or arrays of JJs, and ζ relates to the input effect on the nonlinearity. This means that the small nonlinearity of a JJ can be fully compensated by increasing the driver power. The subsequent analysis focuses on quantum fluctuations and the gain related to small signals. However, Eq. 7 can also be used to calculate the DC point gain (reflection coefficient in a classical sense) $\alpha_{out}/\alpha_{in}$ [7]. Finally, the linearized response for weak quantum signals can be expressed as:

$$\dot{\delta a} = \left[ i\left(\omega_p - \omega_{eff}\right) - \frac{\kappa}{2} - i2K\,|\,\alpha\,|^2 \right]\delta a - iK\delta a^{+}\alpha^{2} + \sqrt{\kappa}\,\delta a_{in} \tag{8}$$

Since the equation derived and its conjugate is linear, so one can transform it to the Fourier domain and decompose all the modes relevant as:

$$-i\left(\omega_s - \omega_p\right)\delta a = i\left[ \frac{\left(\omega_p - \omega_{eff}\right)}{\kappa} + i\frac{1}{2} - 2\frac{K}{\kappa}\,|\,\alpha\,|^2 \right]\delta a - i\frac{K}{\kappa}\delta a^{+}\alpha^{2} + \frac{\sqrt{\kappa}}{\kappa}\delta a_{in}$$

$$\xrightarrow{\Delta = \frac{\omega_s - \omega_p}{\kappa}} 0 = i\left[ \delta + \Delta + \frac{i}{2} - 2\zeta n \right]\delta a - i\zeta n e^{(2i\varphi)}\delta a^{+} + \delta \hat{\alpha}_{in} \tag{9}$$

where Δ is the signal detuning. The last step to calculate the gain is using the conjugate of the Eq. 9 and establish the relating scattering matrix, then employing the input-output formula, the gain of the signal and idler can be calculated [7, 19, 20]. It is shown as follows:

$$\begin{cases} \left[ i\left(-\delta - \Delta + 2\zeta n\right) + \frac{1}{2} \right]\delta a_{\omega} + i\zeta n e^{(2i\varphi)}\delta a_{\omega}^{+} = \delta \hat{\alpha}_{in\omega} \\ \left[ i\left(\delta - \Delta - 2\zeta n\right) + \frac{1}{2} \right]\delta a_{\omega}^{+} - i\zeta n e^{(-2i\varphi)}\delta a_{\omega} = \delta \hat{\alpha}_{in\omega}^{+} \end{cases}$$

$$\longrightarrow \begin{bmatrix} i\left(-\delta - \Delta + 2\zeta n\right) + 0.5 & i\zeta n e^{(2i\varphi)} \\ -i\zeta n e^{(-2i\varphi)} & i\left(\delta - \Delta - 2\zeta n\right) + 0.5 \end{bmatrix} \begin{bmatrix} \delta a_{\omega} \\ \delta a_{\omega}^{+} \end{bmatrix} = \begin{bmatrix} \delta \hat{\alpha}_{in\omega} \\ \delta \hat{\alpha}_{in\omega}^{+} \end{bmatrix} \tag{10}$$

Finally using a few algebras, the BJPA gain is calculated as:

$$G_{BJPA} = \left\{ \frac{i\sqrt{\kappa}\left(\delta - \Delta - 2\zeta n\right) + 0.5}{\left[ -\left\{\left(\delta - \Delta - 2\zeta n\right) + 0.5\right\}\left\{\left(-\delta - \Delta + 2\zeta n\right) + 0.5\right\} - \zeta^2 n^2 \right]} - 1 \right\} - \left\{ \frac{i\sqrt{\kappa}\,\zeta n e^{(2i\varphi)}}{\left[ -\left\{\left(\delta - \Delta - 2\zeta n\right) + 0.5\right\}\left\{\left(-\delta - \Delta + 2\zeta n\right) + 0.5\right\} - \zeta^2 n^2 \right]} \right\} \tag{11}$$

where the first term contributes to the signal gain and the second terms is the idler gain. It is clear from the gain relationship in Eq. 11, that it is dramatically affected by some quantities such as signal and pump detuning, the relative strength of the nonlinearity, and also input coupling rate.

**Results and Discussions**

One of the key innovations of the BJPA is its ability to control nonlinearity, introducing several unique properties. We illustrate this ability with simulations. First, the normalized intra-cavity average number of photons (n), which significantly affects the amplifier's gain, is analyzed for a typical Blochnium with N = 70 and M = 16, as shown in Fig. 2a. Equation 7 highlights the strong dependency of n on δ and ζ, which is evident in the figure shown. Additionally, increasing ζ, particularly when δ ≠ 0, can lead to bifurcation [7], in which the amplifier enters to the nonlinear region. It should be noted that the parametric amplifiers designed in this study operate well away from the bifurcation region to address nonlinearity issues and improve the JPA's compression point. Fig. 2b shows the signal gain of the amplifier as the relative strength of the nonlinearity increases. The increase in signal gain does not fully align with the intra-cavity average number of photons because the signal gain depends on other factors besides n as presented in Equation 11. This discrepancy may be due to the nonlinear factor of $n^2\zeta^2$ in the gain relationship, where increasing ζ leads to higher gain when pump detuning $\delta \neq 0$. One of the remarkable degrees of freedom that the Blochnium structure offers the designers is the ability to adjust M and N simultaneously to enhance the JPA's characteristics, such as gain and linearity. As discussed in the previous section on quantum mechanics, modifying the number of quartons and slave SQUIDs can significantly alter the device's nonlinearity. The results shown in Fig. 2c indicate that while the number of quartons and slave SQUIDs is increasing, it is primarily the number of SQUIDs that has a substantial impact on the JPA's gain. It is due to the fact that, the basic operational phenomenon of JPAs relies on the parametric interaction, where the nonlinearity provided by the SQUIDs is essential for amplifying the signal. Since the SQUIDs are directly responsible for this nonlinearity, the theoretical finding of this study indicates that increasing M has a more profound effect on the amplification process.

To further investigate the signal gain, we simulated the gain versus pump power and signal detuning, illustrated in Fig. 2d. It is clear that the maximum gain occurs around $\delta \neq 0$ and $\Delta = 0$.

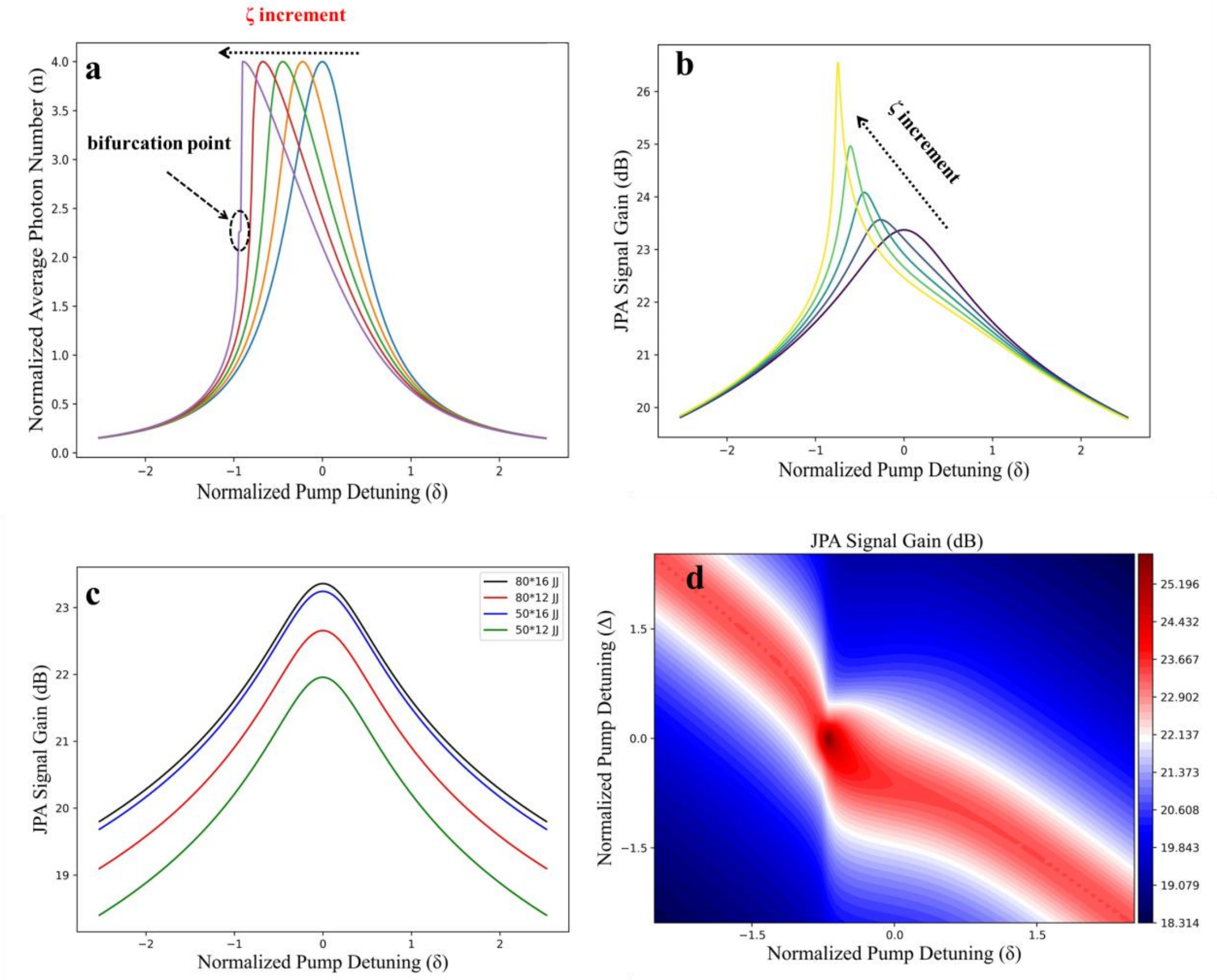

Fig. 2 a) the effect of the relative strength of the nonlinearity on the normalized average photon number (n) vs normalized pump detuning, b) BJPA signal gain vs normalized pump detuning for N = 70 and M = 16, with $\zeta = \zeta = \zeta_0$ *{0.01, 0.1, 0.4, 0.8, 0.95}, $\zeta_0 = -1/\sqrt{27}$ [7], c) BJPA signal gain vs normalized pump detuning for different M and N, d) 2D graph of BJPA signal gain vs normalized pump detuning and signal detuning; BJPA contains N = 70 and M = 16, and $\zeta = 0.01\zeta_0$, $\alpha_c = 0.1$.

It is important to note that adjusting signal gain using a few key parameters is a common practice in other JPA configurations, such as single Josephson Junction (JJ) JPAs or various JJ array topologies [6, 7, 9].

However, the new design goes beyond merely enhancing the amplifier's gain; it primarily aims to significantly improve the amplifier's linearity compared to traditional designs.

The study highlights the significant advantages of BJPA, particularly in terms of their compression point ($P_{1dB}$), which determines the amplifier's linearity limit, and their tunability across the operational bands, like C-band. Enhanced linearity, which is crucial for quantum applications, is achieved through effective management of nonlinearity. The simulation results for compression point of various BJPA are shown in Fig. 3a, revealing that variations in M, N, and $\alpha_c$ significantly impact $P_{1dB}$. This is due to the manipulation of nonlinearity through the Quarton effect introduced in Equation 5. The graph indicates that increasing M (the number of slave JJs in each Quarton) is more effective in improving the compression point compared to increasing N (the number of Quartons). This may be attributed to the critical relationship between M and $\alpha_c$ as described in Equation 5. Simulation results also show that BJPAs can achieve signal gains greater than 25 dB and compression points better than -90 dBm. The corresponding point of $P_{1dB}$ for a structure with N = 70 and M = 8 is marked with dashed lines in Fig. 3a. It is apparent that the compression point achieved here is significantly better with respect to what is reported in [9] for a straight array of Josephson Junctions. This achievement, using a significantly lower number of JJs compared to array structures, simplifies the fabrication process, making it more efficient and manageable.

Simulation results depicted in Fig. 3b, are obtained for a structure with N = 40 and M = 16. The graph illustrates how simply altering φ, the phase of the Josephson junction, manipulates the JJ equivalent inductance ($L_J = L_{J0}/\cos\varphi$), and thereby controlling the effective resonance frequency. This ability enables dynamically adjusting the resonance frequency which allows reaching to an optimized performance across the entire operational bandwidth. Moreover, the parameter $\alpha_c^*$, which represents the ratio between the Josephson energy of the master and slave in a Quarton, also influences the key features of JPA, such as the compression point ($P_{1dB}$). As shown in Fig. 3c, $\alpha_c$ directly affects the nonlinearity achieved by a typical Blochnium structure. By manipulating $\alpha_c$, one can effectively adjust the compression point of the JPA. The graph shows that increasing $\alpha_c$ significantly improves the compression point. To understand this contribution, consider the definition of K as the device nonlinearity factor in Eq. 6 and the substantial impact of $\alpha_c$ on the defined nonlinearity factor. Finally, Fig. 3d represents a comparison of JPA linearity between an array of 800 JJ and a BJPA with N = 50 and M = 16. This specific comparison was chosen because JPAs with single array of JJs have been thoroughly explored in both theoretical and experimental studies [5,6, 9]. The results indicate that while the BJPA achieves an appropriate gain comparable to the JPAs discussed in those references [5,6, 9], it offers a significantly improved compression point. Although the proposed structure enhances $P_{1dB}$ dramatically and exhibits an impressive capability to cover the C-band, an error margin should be considered. The deviation mentioned may arise from factors such as fabrication inaccuracies, variations in material properties, and environmental fluctuations, all of which can affect the consistency and reliability of the amplification process. Nevertheless, the comparison with existing theoretical and experimental studies for the previously reported structures [5,6, 9] suggests that our simulation results are promising. In addition to the features discussed, a BJPA offers several practical advantages; its design facilitates better integration having much fewer elements holding smaller space and more reliable operation, enabling seamless incorporation into current quantum computing architectures [10]. Interestingly, if the BJPA can surpass $P_{1dB}$ of -90 dBm, which is confirmed by the theoretical results, it opens the possibility of using a two-stage JPA configuration instead of a single-stage for amplifying quantum signals. This approach would eliminate the need for the additional amplifiers such as HEMT [24, 25], thereby avoiding the introduction of extra noise into the quantum signals and increasing the integrity of the circuit designed.

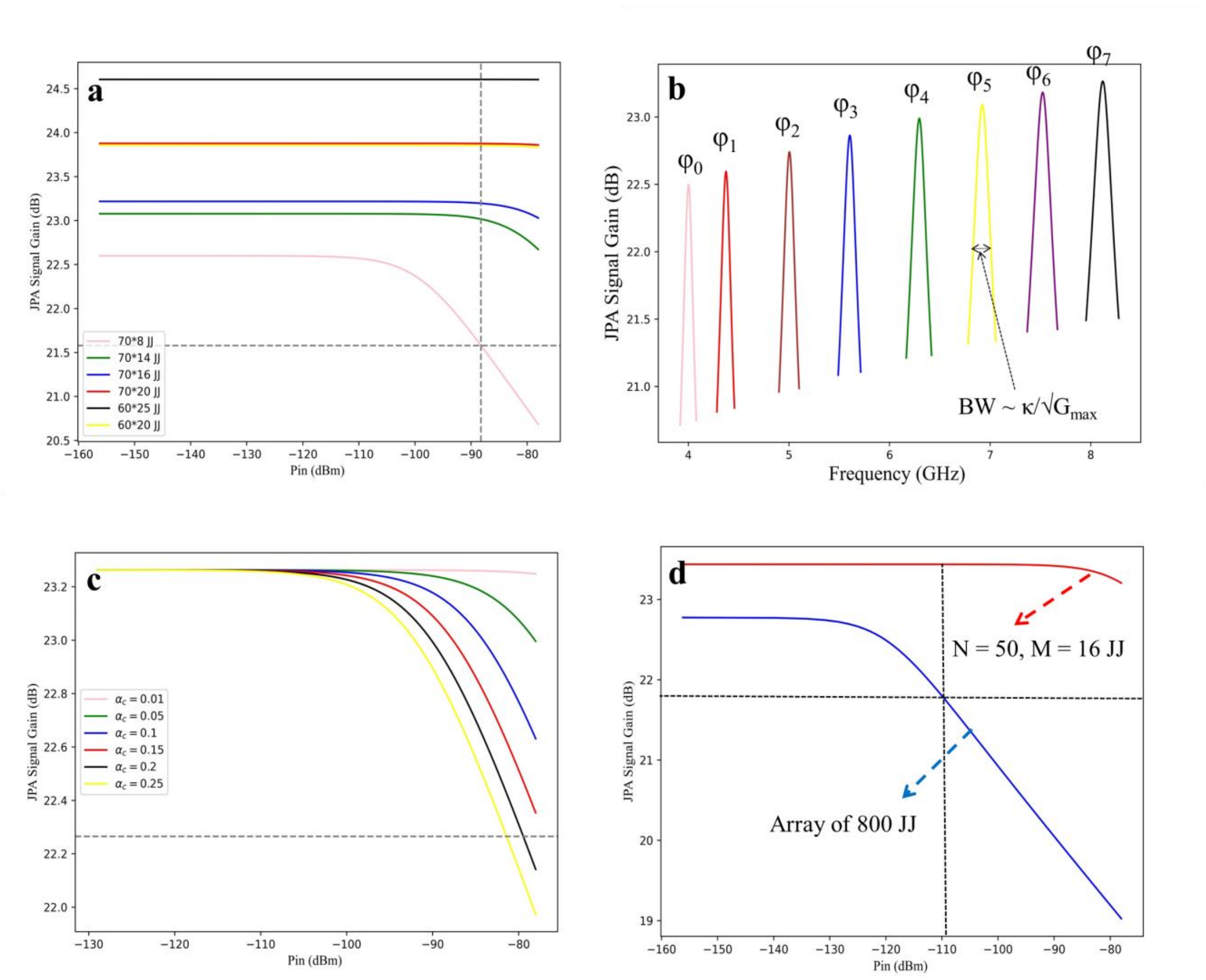

Fig. 3 a) the effect of the number of quarton and slave SQUIDs on the compression points $P_{1dB}$ vs the pump power, $P_{in}$ (dBm), $\alpha_c = 0.1$, b) the effect of the JJ's nonlinear inductor on the BJPA's signal gain and sweeping throughout the C-band for N = 40 and M = 14, $\alpha_c = 0.1$, BW: bandwidth, c) the effect of the nonlinearity factor $\alpha_c$ on $P_{1dB}$ for N = 40 and M = 14, d) the comparison of the JPA linearity between a straight array of 800 JJ and a BJPA with N = 50 and M = 16; For all figures the signal and pump detuning are considered to be zero $\Delta = \delta = 0$, and $\zeta = 0.01\zeta_0$.

The enhanced tunability and control over operational parameters provide researchers and engineers with a versatile tool for optimizing system performance. This adaptability is particularly valuable in experimental setups where precise control over qubit interactions and readout processes is fundamentally important. Furthermore, the BJPA superior performance characteristics, such as high signal gain and improved compression point, contribute to the overall stability and reliability of quantum systems. The ability to finely tune the BJPA properties also opens up new avenues for exploring novel quantum phenomena and enhancing the fidelity of quantum operations. The findings of this study highlight the potential of BJPA to significantly advance the field of quantum technology, providing a robust platform for future innovations in scalable quantum systems.

## Conclusions

In this work, we attempted to design a new structure as a high performance JPA to handle some limitations of the traditional JPA such a limited and small compression point ($P_{1dB}$). We also introduced a tunable JPA structure to sweep over the operational frequency band of superconducting qubits. For this purpose, the named Blochnium structure was defined and theoretically its formulism was derived including its contributed Hamiltonian, and finally the related dynamic equation of motion. Using the quantum theory, the effect of the Blochnium structure on the JJ nonlinearity is analyzed completely. As a result, the BJPA stands out due to its advanced control over nonlinearity, which introduces several unique properties essential for quantum computing applications. Its ability to finely tune parameters such as N, M, and φ allows for precise management of gain and compression points, facilitating enhanced performance across the entire C-band. The capability to sweep this band without significant spectral overlap is particularly beneficial for frequency multiplexing in scalable quantum systems, reducing quantum errors and improving system reliability. The combination of high signal gain, superior compression point, and narrow instantaneous bandwidth underscores the BJPA potential as a powerful tool in quantum technology. Furthermore, the BJPA design ensures seamless integration with existing superconducting circuit technology, offering researchers and engineers a versatile and adaptable platform for optimizing quantum system performance. These attributes, along with the Blochnium ability to explore novel quantum phenomena, position it as a promising candidate for future innovations in scalable quantum systems, significantly advancing the field of quantum technology.